\documentclass[aps,prl,twocolumn,superscriptaddress,showpacs,preprintnumbers,amsmath,amssymb]{revtex4}

\usepackage{graphicx}
\usepackage{dcolumn}
\usepackage{bm}
\usepackage{xspace}
\usepackage{color}

\newcommand{\ups}{\ensuremath{\Upsilon (4S)}\xspace}
\newcommand{\B}{\ensuremath{B}\xspace}
\newcommand{\bz}{\ensuremath{\B^0}\xspace}

\newcommand{\bzb}{\ensuremath{\overline{B}{}^0}\xspace}
\newcommand{\bzbzb}{\ensuremath{\bz\bzb}\xspace}

\newcommand{\bzbz}{\ensuremath{\bz\bz}\xspace}
\newcommand{\bzbbzb}{\ensuremath{\bzb\bzb}\xspace}
\newcommand{\bbb}{\ensuremath{\B\overline{B}}\xspace}

\newcommand{\decay}[2]{\ensuremath{#1\to #2}\xspace}

\newcommand{\dt}{\ensuremath{\Delta t}\xspace}
\newcommand{\dm}{\ensuremath{\Delta m_d}\xspace}

\newcommand{\dgammar}{\ensuremath{\frac{\Delta \Gamma_d}{\Gamma_d}}\xspace}
\newcommand{\tmin}{\ensuremath{t_\mathrm{min}}\xspace}

\newcommand{\SF}{\ensuremath{\mathrm{SF}}\xspace}
\newcommand{\OF}{\ensuremath{\mathrm{OF}}\xspace}
\newcommand{\QM}{\ensuremath{\mathrm{QM}}\xspace}
\newcommand{\SD}{\ensuremath{\mathrm{SD}}\xspace}

\newcommand{\APSmax}{\ensuremath{\mathrm{A_{PS}^{\rm max}} }\xspace}
\newcommand{\APSmin}{\ensuremath{\mathrm{A_{PS}^{\rm min}} }\xspace}

\begin{document}
version 7.6
\title{
 Measurement of EPR-type flavour entanglement in \decay{\ups}{\bzbzb} decays}
\affiliation{Budker Institute of Nuclear Physics, Novosibirsk}
\affiliation{University of Cincinnati, Cincinnati, Ohio 45221}
\affiliation{Justus-Liebig-Universit\"at Gie\ss{}en, Gie\ss{}en}
\affiliation{The Graduate University for Advanced Studies, Hayama}
\affiliation{Hanyang University, Seoul}
\affiliation{University of Hawaii, Honolulu, Hawaii 96822}
\affiliation{High Energy Accelerator Research Organization (KEK), Tsukuba}
\affiliation{University of Illinois at Urbana-Champaign, Urbana, Illinois 61801}
\affiliation{Institute of High Energy Physics, Chinese Academy of Sciences, Beijing}
\affiliation{Institute of High Energy Physics, Vienna}
\affiliation{Institute of High Energy Physics, Protvino}
\affiliation{Institute for Theoretical and Experimental Physics, Moscow}
\affiliation{J. Stefan Institute, Ljubljana}
\affiliation{Korea University, Seoul}
\affiliation{Kyungpook National University, Taegu}
\affiliation{Ecole Polytechnique F\'ed\'erale Lausanne, Lausanne}
\affiliation{University of Ljubljana, Ljubljana}
\affiliation{University of Maribor, Maribor}
\affiliation{University of Melbourne, Victoria}
\affiliation{Nagoya University, Nagoya}
\affiliation{Nara Women's University, Nara}
\affiliation{National Central University, Chung-li}
\affiliation{National United University, Miao Li}
\affiliation{Department of Physics, National Taiwan University, Taipei}
\affiliation{H. Niewodniczanski Institute of Nuclear Physics, Krakow}
\affiliation{Nippon Dental University, Niigata}
\affiliation{Niigata University, Niigata}
\affiliation{University of Nova Gorica, Nova Gorica}
\affiliation{Osaka City University, Osaka}
\affiliation{Osaka University, Osaka}
\affiliation{Panjab University, Chandigarh}
\affiliation{Peking University, Beijing}
\affiliation{RIKEN BNL Research Center, Upton, New York 11973}
\affiliation{University of Science and Technology of China, Hefei}
\affiliation{Seoul National University, Seoul}
\affiliation{Shinshu University, Nagano}
\affiliation{Sungkyunkwan University, Suwon}
\affiliation{University of Sydney, Sydney NSW}
\affiliation{Tata Institute of Fundamental Research, Mumbai}
\affiliation{Toho University, Funabashi}
\affiliation{Tohoku Gakuin University, Tagajo}
\affiliation{Tohoku University, Sendai}
\affiliation{Department of Physics, University of Tokyo, Tokyo}
\affiliation{Tokyo Institute of Technology, Tokyo}
\affiliation{Tokyo Metropolitan University, Tokyo}
\affiliation{Tokyo University of Agriculture and Technology, Tokyo}
\affiliation{Virginia Polytechnic Institute and State University, Blacksburg, Virginia 24061}
\affiliation{Yonsei University, Seoul}
\author{A.~Go}\affiliation{National Central University, Chung-li} 
\author{A.~Bay}\affiliation{Ecole Polytechnique F\'ed\'erale Lausanne, Lausanne} 
\author{K.~Abe}\affiliation{High Energy Accelerator Research Organization (KEK), Tsukuba} 
  \author{H.~Aihara}\affiliation{Department of Physics, University of Tokyo, Tokyo} 
  \author{D.~Anipko}\affiliation{Budker Institute of Nuclear Physics, Novosibirsk} 
  \author{V.~Aulchenko}\affiliation{Budker Institute of Nuclear Physics, Novosibirsk} 
  \author{T.~Aushev}\affiliation{Ecole Polytechnique F\'ed\'erale Lausanne, Lausanne}\affiliation{Institute for Theoretical and Experimental Physics, Moscow} 
  \author{A.~M.~Bakich}\affiliation{University of Sydney, Sydney NSW} 
  \author{E.~Barberio}\affiliation{University of Melbourne, Victoria} 
%
  \author{K.~Belous}\affiliation{Institute of High Energy Physics, Protvino} 
% ...
\collaboration{The Belle Collaboration}


\date{\today}

\begin{abstract}
The neutral \B-meson pair produced at the \ups should exhibit 
a non-local correlation of the type discussed by Einstein, Podolski, and Rosen.
We measure this correlation using the time-dependent flavour asymmetry of
semileptonic \bz decays, which we compare with predictions from quantum
mechanics and two local realistic models. The data are consistent with
quantum mechanics, and inconsistent with the other models.
Assuming that some \B pairs disentangle to produce
\bz and \bzb with definite flavour,
we find a decoherent fraction of $0.029\pm0.057$,
consistent with no decoherence.
\end{abstract}


\maketitle

The concept of entangled states, which cannot be described
as product states of their parts, was born 
with Quantum Mechanics (QM).
In 1935 Einstein, Podolski and
Rosen (EPR) considered such a pair of particles 
and concluded that QM cannot
be a ``complete'' theory~\cite{EPR}; 
this suggests that additional (``hidden'') variables are required. 
In 1964 J.~S.~Bell showed that
QM can violate a certain inequality, which is (by contrast) satisfied
by all local hidden variable models~\cite{Bell}.
%
{Many experiments have since been performed and found excellent agreement
with the prediction of QM (although no ``loophole''-free 
Bell test has yet been performed)~\cite{BellExp}. Most of these studies have
used pairs of optical photons; 
it is also interesting to test EPR correlations in massive 
systems~\cite{Bertlmann06} at much higher energies~\cite{Bernabeu06}.} 
In this Letter, we present a study of EPR correlation 
in the flavour of \B-meson pairs produced at the \ups.
 {
Contrary to the analysis presented in~\cite{Garda}, and as discussed in
the literature~\cite{Bertlmann}, a Bell inequality test cannot be performed
in this system } 
due to the rapid decrease in time of the \B-meson amplitudes,
and the passive character of the flavour measurement,
\emph{via} reconstruction of \B-meson decay products.
 {
Instead, we compare the data with predictions from QM and 
other models. Related studies have been performed in the
$K$-meson system~\cite{CPLEAR-KLOE,b-g-h}
to test decoherence~\cite{decoherence} effects;
$\ups \to \bz\bzb$ data have also been analyzed,
but using time-integrated information only~\cite{bmes-decoherence}.
Here, we use information on reconstructed $B$-meson decay times
to test both decoherence and the Pompili-Selleri
model~\cite{Selleri}, which represents a range of possible
local hidden-variable theories~\cite{Santos}. 
} 

The wavefunction of a \bzbzb pair from \ups decay is
analogous to that of photons in a spin-singlet state 
~\cite{B0B0bar,
  GisinGo}:
\begin{equation}
|\psi\rangle = \frac{1}{\sqrt{2}} 
        \left[\left|\bz \right\rangle_1\otimes  \left|\bzb \right\rangle_2 -
              \left|\bzb \right\rangle_1\otimes \left|\bz  \right\rangle_2\right].
\label{B0B0b}
\end{equation}
Decays occurring at the same proper time are fully correlated:
the flavour-specific decay of one meson
fixes the (previously undetermined) flavour \bz/\bzb 
of the other meson.
Given (\ref{B0B0b}), the time-dependent rate for decay 
into two flavour-specific states
$R_i = e^{-\Delta t/\tau_{B^0}}/(4\tau_{B^0})\{ 1 \pm \cos(\Delta m_d \Delta t) \}$
for
opposite flavour (\bzbzb; $+$, $i=\mathrm{OF}$) and
same flavour (\bzbz or \bzbbzb; $-$, $i=\mathrm{SF}$) decays.
$\Delta t \equiv |t_1 - t_2|$ is the proper-time difference 
of the decays, and $\Delta m_d$ the mass difference
between the two \bz-\bzb mass eigenstates.
We have assumed a lifetime difference $\Delta \Gamma_d = 0$
and neglected the effects of $CP$ violation in mixing, which  are $O(10^{-4})$ or less.

 {Thus in QM} the time-dependent asymmetry 
 {$A(\dt) \equiv \left(R_\OF-R_\SF\right)/\left(R_\OF+R_\SF\right)
= \cos(\dm\dt)$,} 
is a function of $\Delta t$ but not the individual times $t_{1,2}$.
 {This is} a manifestation of entanglement.
%
\begin{figure}[htb] 
\includegraphics[height=4.5cm]{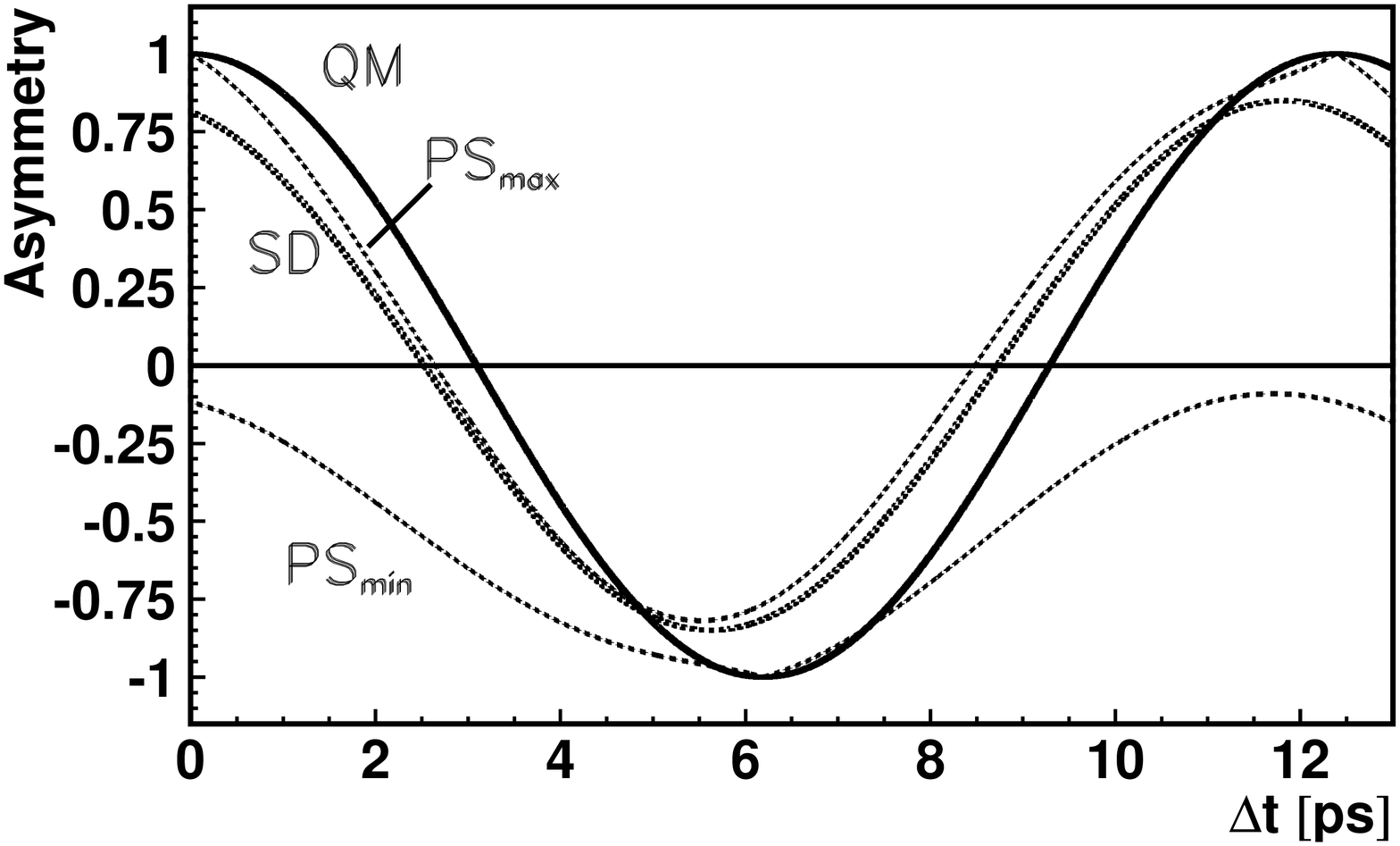}
\caption{Time-dependent asymmetry predicted by 
(QM) quantum mechanics
and
(SD) spontaneous and immediate disentanglement of the \B-pair~\cite{Furry,decoherence},
and 
(PS$_{min}$ to PS$_{max}$) the range of asymmetries allowed by the Pompili-Selleri model~\cite{Selleri}.
$\Delta m_d = 0.507\,\mathrm{ps}^{-1}$ is assumed~\cite{PDG06}.}
\label{fig:models}
\end{figure}
%
 {By contrast, we can consider Spontaneous Disentanglement (SD),
an extreme case of decoherence,
in which the \B-meson pair immediately separates into a \bz and \bzb
with well-defined flavour,} 
which then evolve independently~\cite{Furry}.
 {The asymmetry becomes} 
\begin{eqnarray}
A_{\SD}(t_1,t_2) &=& \cos(\dm t_1)\cos(\dm t_2)  \\
       &=& \frac{1}{2}[\cos(\dm (t_1+t_2))+\cos(\dm \Delta t)], \nonumber
\end{eqnarray}
depending on $t_1+t_2$ in addition to \dt.
Because of the large uncertainty on the \ups decay point,
it is difficult to measure individual decay times $t_{1,2}$:
only \dt is measured in this analysis.
 {
If we first integrate the OF and SF distributions keeping \dt constant we obtain the asymmetry}
curve shown in Fig.~\ref{fig:models},
 {which differs significantly from the simple cosine term due to QM
(also shown).} 

In the  {model of} Pompili and Selleri (PS)~\cite{Selleri}, each \B has
well-defined 
flavour, $B^0$ or $\overline{B}{}^0$, and mass,
corresponding to the heavy and light \bz-\bzb eigenstates.
There are thus four basic states: $B^0_H$, $B^0_L$, $\overline{B}{}^0_H$, $\overline{B}{}^0_L$.
 {
  At equal times $\dt=0$,
  the $B$-mesons in a pair have opposite values of both mass and flavour;
  mass values are stable, but the flavour can change, simultaneously for the 
  two mesons.
  There are no other assumptions, except a requirement 
  that QM predictions for uncorrelated $B$-decays are reproduced.
  This rather general scheme includes a range of possible local realistic models,
  and allows time-dependent asymmetries to lie within the bounds
}
\begin{align}
\APSmax(t_1,t_2)& = 1-|\{1-\cos(\dm \Delta t)\}\cos(\dm \tmin) \nonumber\\
		& +\sin(\dm \Delta t)\sin(\dm \tmin)|,\;\text{and}
		\label{eq-APSmax}					\\
\APSmin(t_1,t_2)& = 1-\min(2+\Psi,\,2-\Psi),\;\text{where}
		\label{eq-APSmin}					\\
\Psi		& = \{1+\cos(\dm \Delta t)\}\cos(\dm \tmin) \nonumber	\\
		& - \sin(\dm \Delta t)\sin(\dm \tmin).
\end{align}
Note the additional $\tmin=\mathrm{min}(t_1,t_2)$ dependence.
 {
After integration for fixed values of \dt
}
we obtain the  {asymmetry} curves $\rm PS_{max}$ and $\rm PS_{min}$
shown in Fig.~\ref{fig:models}.

To determine the asymmetry,
we use $152\times 10^6$ \bbb pairs collected by the Belle detector at the \ups
resonance at the KEKB asymmetric-energy
(3.5 GeV on 8.0~GeV) $e^+e^-$ collider \cite{KEKB}. The Belle detector~\cite{BELLE}
is a large-solid-angle spectrometer 
consisting of a silicon vertex detector (SVD),
central drift chamber (CDC), aerogel Cherenkov
counters (ACC), time-of-flight
counters (TOF), and a CsI(Tl) electromagnetic calorimeter
(ECL) inside a 1.5T superconducting
solenoid. The flux return
is instrumented to detect $K^0_L$ and
identify muons (KLM). The \ups is produced with 
$\beta\gamma = 0.425$ close to the $z$ axis (defined as anti-parallel to the positron beam line). As
the \B momentum is low in the \ups
center-of-mass system (CMS), \dt can be determined from 
the $z$-displacement of \B-decay vertices: $\dt
\approx \Delta z/\beta\gamma c$.

We use an event selection similar to that of
a previous Belle analysis~\cite{KojiHara,physrevd71}, 
but optimised for theoretical model discrimination;
in particular we use more stringent criteria
on the flavour tag purity than the previous analysis.
To enable direct comparison of the result with
different models, we subtract both background and mistagged-flavour events 
from the data, and then correct for detector effects
by deconvolution.

We determine the flavour of one neutral \B by reconstructing
the decay \decay{\bz}{D^{*-}\ell^+\nu},
with \decay{D^{*-}}{\overline{D}{}^0 \pi^-_s}
and
$\overline{D}{}^0 \to K^+ \pi^-(\pi^0)$ or
$K^+ \pi^- \pi^+ \pi^-$
(charge-conjugate modes are included throughout this Letter).
Charged particles (except the ``slow pion'' $\pi_s$)
are chosen from tracks with associated SVD hits and 
radial impact parameter $dr<0.2$ cm,
and required to satisfy kaon/pion 
identification criteria using combined 
TOF, ACC and CDC ($dE/dx$) information~\cite{KpiID}.
$\pi^0\to\gamma\gamma$ candidates are
selected with $|M_{\gamma\gamma}-m_{\pi^0}| < 11\,\mathrm{MeV}/c^2$ and
momenta $p_{\pi^0} > 0.2\,\mathrm{GeV}/c$;
the photons 
must have energies $E_\gamma > 80\,\mathrm{MeV}$.
We select $D^0$ candidates 
with $(M_{Kn\pi}-m_{D^0}) \in [-13,13]\,\mathrm{MeV}/c^2$ for
$K\pi(\pi\pi)$ and
$[-37,23]\,\mathrm{MeV}/c^2$ for $K \pi \pi^0$.
A $D^*$ candidate is
formed by constraining a $D^0$ and slow pion (having opposite charge to the lepton) to a common vertex.
We require a mass difference 
$M_{\rm diff}=M_{Kn\pi\pi_s} - M_{Kn\pi}
\in [144.4,146.4]~\mathrm{MeV}/c^2$, and CMS momentum
$p^*_{D^*} < 2.6~\mathrm{GeV}/c$, consistent with 
\B-decay. 
Electron identification uses momentum and $dE/dx$ information, 
ACC response, and energy deposition in the ECL.
Muon identification is based on penetration depth and matching of hits in the KLM to the extrapolated track.
The efficiency is about 92\% (84\%) for electrons (muons) in the relevant momentum region,
from 1.4 to 2.4 GeV$/c$ in the CMS;
hadrons pass this selection with an efficiency of 0.2\% (1.1\%). 
%
%
We require that the CMS angle between the $D^{*}$
and lepton be greater than $90^{\circ}$. From the relation
$M_{\nu}^2=(E^*_B-E^*_{D^*\ell})^2-|\vec{p}^{\,*}_B|^2-|\vec{p}^{\,*}_{D^*\ell}|^2+ 2 |\vec{p}^{\,*}_B| |\vec{p}^{\,*}_{D^*\ell}| \cos(\theta_{B,D^*\ell})$,
where $\theta_{B,D^*\ell}$ is the angle between $\vec{p}^{\,*}_B$ and $\vec{p}^{\,*}_{D^*\ell}$,
we can reconstruct $\cos(\theta_{B,D^*\ell})$ by assuming a vanishing neutrino mass.
We require $|\cos(\theta_{B,D^*\ell})|<1.1$. 
The neutral \B decay position is determined by fitting
the lepton track and $D^0$ trajectory to a vertex, constrained to lie
in the $e^+e^-$ interaction region (smeared in the $r-\phi$ plane to account for the \B
flight length); we require $\chi^2/n_{dof} < 75$.

The remaining tracks are used to
determine the second \B decay vertex and its
flavour, using the method of Ref.~\cite{Btagging,KpiID}.
Events are
classified into six subsets according to the purity of the tag. In this
analysis we use only leptonic tags from the highest purity subset.

In total, 8565 events are selected (6718 OF, 1847 SF).
A GEANT-based Monte Carlo (MC) sample assuming
QM correlation, with five times the number of events, was
analysed with identical criteria;
its $\Delta z$ and $D^*$ mass distributions were tuned to those of
the data. This sample was used for consistency
checks, background estimates and subtraction, and to build deconvolution
matrices.

\begin{table*}
\caption{Time-dependent asymmetry in \dt bins, corrected for experimental effects,
with statistical and systematic uncertainties.
Contributions from event selection,
background subtraction, wrong tag correction, and deconvolution are also shown.
}
\begin{center}
\begin{ruledtabular}
\begin{tabular}{crccccccc}
  &			  &				  &		&  \multicolumn{5}{c}{Systematic errors} \\ \cline{5-9}
$\Delta t$ bin
  & window [ps]           & $A$ and total error           & statistical error
									&     total  & event sel. &bkgd sub.&  wrong tags & deconvolution \\
\hline
 1& 0.0 -- \phantom{2}0.5 &  $\phantom{-}1.013\pm0.028$ &     0.020	&     0.019  & 0.005      & 0.006   &      0.010  &  0.014 \\
 2& 0.5 -- \phantom{2}1.0 &  $\phantom{-}0.916\pm0.022$ &     0.015	&     0.016  & 0.006      & 0.007   &      0.010  &  0.009 \\
 3& 1.0 -- \phantom{2}2.0 &  $\phantom{-}0.699\pm0.038$ &     0.029	&     0.024  & 0.013      & 0.005   &      0.009  &  0.017 \\
 4& 2.0 -- \phantom{2}3.0 &  $\phantom{-}0.339\pm0.056$ &     0.047	&     0.031  & 0.008      & 0.005   &      0.007  &  0.029 \\
 5& 3.0 -- \phantom{2}4.0 &            $-0.136\pm0.075$ &     0.060	&     0.045  & 0.009      & 0.009   &      0.007  &  0.042 \\
 6& 4.0 -- \phantom{2}5.0 &            $-0.634\pm0.084$ &     0.062	&     0.057  & 0.021      & 0.014   &      0.013  &  0.049 \\
 7& 5.0 -- \phantom{2}6.0 &            $-0.961\pm0.077$ &     0.060	&     0.048  & 0.020      & 0.017   &      0.012  &  0.038 \\
 8& 6.0 -- \phantom{2}7.0 &            $-0.974\pm0.080$ &     0.060	&     0.053  & 0.034      & 0.025   &      0.020  &  0.025 \\
 9& 7.0 -- \phantom{2}9.0 &            $-0.675\pm0.109$ &     0.092	&     0.058  & 0.041      & 0.027   &      0.022  &  0.022 \\
10& 9.0 --           13.0 &  $\phantom{-}0.089\pm0.193$ &     0.161	&     0.107  & 0.067      & 0.063   &      0.038  &  0.039 \\
11&13.0 --           20.0 &  $\phantom{-}0.243\pm0.435$ &     0.240	&     0.363  & 0.145      & 0.226   &      0.080  &  0.231 \\
\end{tabular}
\end{ruledtabular}
\end{center}
\label{tab:results}
\end{table*}

To compensate for the rapid fall in event rate with
\dt, the
time-dependent distributions are histogrammed in 11 variable-size
bins (Table~\ref{tab:results}).
Background subtraction is then performed bin-by-bin;
systematic errors are likewise determined
by estimating variations in the OF and SF distributions, and 
calculating the effect on the asymmetry.
Terms due to event selection are estimated by comparing data 
and MC distributions for each quantity, and
converting discrepancies into yield variations: effects due
to each selection are added in quadrature. Estimation of the remaining
terms is described below.
\begin{figure*}[!ht]
\includegraphics[width=0.95\textwidth,height=!,bb=0 0 1701 494,clip]{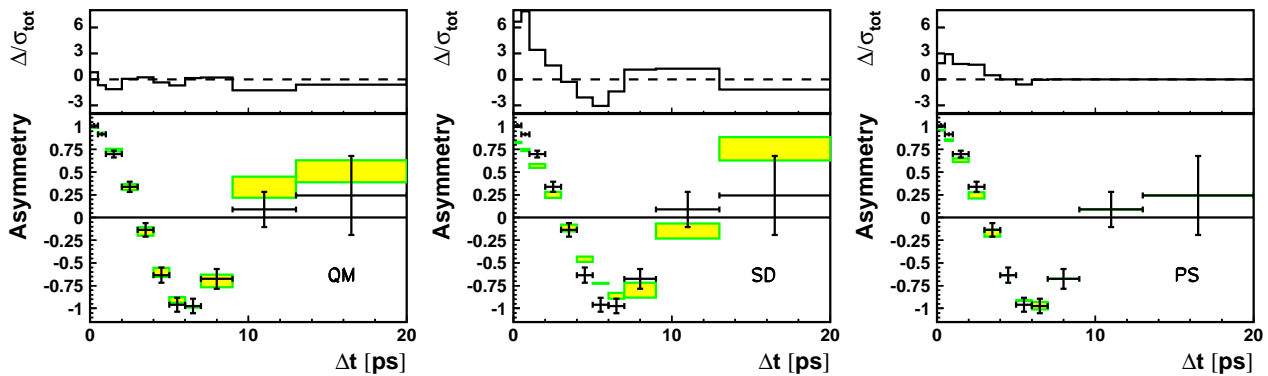}
\caption{
Bottom: time-dependent flavour asymmetry (crosses) 
and the results of weighted least-squares fits to the (left to right) QM,
SD, and PS models (rectangles, showing $\pm 1 \sigma$ errors on \dm).
Top: differences $\Delta \equiv A_{\text{data}}-A_{\text{model}}$ in each bin, divided
by the total experimental error $\sigma_{\text{tot}}$.
Bins where $\APSmin < A_{\text{data}} < \APSmax$ have been assigned a null deviation:
see the text.
}
\label{fig:asym}
\end{figure*}

Four types of background events have been considered: 
$e^+e^- \rightarrow q\bar q$ continuum, non-$D^*$ events, wrong $D^*$--lepton combinations, and $B^+
\rightarrow \overline{D}{}^{**0}\ell\nu$ events. Off-resonance data ($8.3~\rm{fb^{-1}}$)
were used to estimate the continuum background, which was found to be
negligible.

The background to the $D^0$ sample, and misassigned slow pions, produce a
background under the $D^*$ peak in $M_{\rm diff}$. As a correction,
we subtract $126\pm6$ ($54\pm 4$) 
such OF (SF) events based on scaled yields from the sideband
%
$M_{\rm diff} \in [156.0,164.0]\,\mathrm{MeV}/c^2$.
%
The corresponding systematic uncertainty is estimated by considering 
statistical fluctuations, and moving cuts by $\pm 0.02 \ \mathrm{MeV}/c^2$
(the estimated mis-calibration in $M_{\rm diff}$).
Alternate sidebands 
$[152.0,156.0]$ and
$[164.0,168.0]\,\mathrm{MeV}/c^2$ are also used:
the difference from default results (consistent with
statistical fluctuations) is conservatively included
in the systematic error.

The wrong $D^*$--lepton combination background is mainly due to the
combination of a $D^*$ from one \B with a true lepton 
from the other \B, with a smaller fraction due to misidentified leptons,
and from charm decay.
To estimate this background, for each selected lepton which forms a CMS angle 
to the $D^*$ less than $90^{\circ}$, we reverse its CMS momentum
labelling the modified lepton $\ell'$, and require $|\cos(\theta_{B,D^*\ell'})|<1.1$. 
This procedure, intended to reject correlated $D^*\ell$
pairs while selecting events with no angular correlation, has been
validated on MC events where true $B^0 \to D^{*-} \ell^+ X \nu$ combinations
have been excluded. (The correlated background from charm decays is negligible.)
We obtain $78\pm9$ OF and
$237\pm15$ SF events, which are then subtracted.
Contributions to the systematic error are obtained by considering the
statistical fluctuations and by moving cuts by $\pm 0.1$ to account 
for possible data-MC discrepancies.

After these subtractions, three main types of events remain:
$B^0 \rightarrow D^{*-}\ell^+\nu$, the signal;
$B^0 \rightarrow D^{**-}\ell^{+}\nu$, which we retain because it undergoes mixing; and
$B^{+} \rightarrow \overline{D}{}^{**0}\ell^{+}\nu$ background.
MC shapes for the signal and the sum of the $D^{**}$ channels
are used in a two-parameter fit
to the $\cos(\theta_{B,D^*\ell})$ distribution
to find the total $D^{**}$ contribution
($\chi^2/n_{dof}=56/46$),
and its $B^{+}$ component is 
then estimated using MC fractions. We find $255.5\pm16.0$ events
(254.0 OF and 1.5 SF),
which we subtract from the data. The systematic uncertainty is
estimated by adding in quadrature the fit error (6\%) and variations
obtained by moving the fit region (3\%) and changing to a
single parameter fit with forced normalisation (2\%). 
We also assign a 20\% uncertainty on the ratio of branching fractions of
\decay{\bz}{D^{**-}\ell^+\nu} to \decay{B^{+}}{\overline{D}{}^{**0}\ell^+\nu}.

We correct for wrong flavour assignments 
using OF and SF distributions from wrongly-tagged MC events.
The mistag fraction $0.015\pm0.001$ (stat) is consistent with that in data~\cite{physrevd71};
we assign a systematic error of $\pm 0.005$.

Remaining reconstruction effects (e.g.\ resolution in $\Delta t$,
selection efficiency) are corrected by deconvolution,
treating the SF and OF distributions separately. The method is
based on Deconvolution with Singular Value Decomposition
(DSVD)~\cite{SVD};
$11 \times 11$ response matrices are built
separately for SF and OF events, using MC $D^*\ell\nu$ events
indexed by generated and reconstructed \dt values.
The procedure has been optimised by a Toy Monte Carlo (TMC)
technique where sets of
several hundred simulated experiments are generated with data and
MC samples identical in size to those of the real experiment, but
assuming different true asymmetries $A_\QM$, $A_\SD$, and \APSmax.
In particular the following points have been studied:

(1) The effective matrix rank was reduced from 11 to 5 (6) for the OF (SF)
sample, to minimize the total error.
(The statistical precision of some singular values is poor.)

(2) The MC events used to fill the response matrix,
and provide an \emph{a priori} to the regularization algorithm,
introduce a potential bias:
\emph{e.g.}\ the first $\Delta t$ bin contains few SF events
for QM, but is well-populated for SD.
We therefore replace SF and OF samples with mixtures 
$\rm SF + o \times OF$
and $\rm OF + s \times SF $,
choosing $\rm s=o=0.2$ to minimize systematic effects;
the exact values are not critical.

(3)
After DSVD, measured differences from input values
are averaged (over QM, SD, and PS) and subtracted bin-by-bin
from the asymmetry, 
to reduce the potential bias against any one model.
The maximal
absolute deviation of the corrected distribution from the
three models is assigned as the systematic error
in each \dt bin.

(4) 
A $46\,\mu\mathrm{m}$ Gaussian smearing term,
inferred from the difference between MC and data vertex-fit errors,
is used to tune the MC $\Delta z$ distribution to the data.
(The average $\Delta z$ resolution is $\approx 100\,\mu\mathrm{m}$).
This term was varied by its $\pm35\,\mu\mathrm{m}$ uncertainty,
and the resulting bin-by-bin difference in the asymmetry taken as the 
systematic error.

Terms from (3) and (4) are added in quadrature to give the
total systematic error due to deconvolution.
We test the consistency of the method by fitting the \bz decay time
distribution (summing OF and SF samples),
leaving the \bz lifetime as a free parameter. We obtain $1.532 \pm
0.017$(stat)~ps, consistent with the world average~\cite{PDG06}.
We also repeat the deconvolution
procedure using events with better
vertex fit quality, and hence more precise \dt values:
consistent results are obtained.

The final results, which may be directly compared with theoretical models,
are shown in Table~\ref{tab:results};
addition in quadrature is used to combine the various error terms. 

We perform weighted least-squares fits to $A(\Delta t)$,
including a term taking the world-average \dm into account.
To avoid bias
we discard BaBar and Belle measurements,
which assume QM correlations:
this yields $\langle \dm \rangle = (0.496 \pm 0.014)\,\mathrm{ps}^{-1}$~\cite{HFAG}.

In fits to the QM, SD, and PS predictions,
  we obtain
  $\dm = 0.501\pm0.009$, $0.419\pm0.008$, and $0.447\pm0.010$ ps$^{-1}$
  with  $\chi^2$ of 5.2, 174, and 31.3 respectively, for eleven degrees of freedom:
see Fig.~\ref{fig:asym}.
The data favour QM over the SD model at $13\sigma$,
and QM over the PS model at $5.1\sigma$~\cite{FIT}.
 As noted above, $CP$ violation in mixing can be neglected.
Introducing a lifetime difference 
$\dgammar = 0.009 \pm 0.037$~\cite{HFAG} has a negligible effect on
the fit.
As a consistency check, the time-dependent asymmetry before deconvolution
is compared to MC predictions for QM and (\emph{via} reweighting) the SD and PS models:
QM is strongly favoured.

 {
Following other phenomenological studies of decoherence (\emph{e.g.}\ Ref.~\cite{b-g-h})
we also fit the data with the function $(1-\zeta_{\bz\bzb})A_\QM + \zeta_{\bz\bzb} A_\SD$:
this is equivalent to modifying the interference term in the \bz-\bzb\ basis, 
or to} assuming that only a fraction of
the neutral \B pairs from \ups decays disentangle immediately into a
\bz and a \bzb. 
 {We find} $\zeta_{\bz\bzb} =0.029\pm0.057$,
consistent with no decoherence.

In summary, we have analysed neutral \B pairs produced by \ups decay,
determined the time-dependent asymmetry due to flavour oscillations,
and corrected for experimental effects by 
deconvolution: the results
can be directly compared to theoretical
models. 
 {Any local realistic model including the assumptions} of Pompili and
Selleri is strongly disfavoured compared  {to quantum mechanics.} 
Immediate disentanglement, in which
definite-flavour \bz and \bzb evolve independently, is ruled out; 
if a fraction of \B-pairs is assumed to decay incoherently,
we find a decoherent fraction consistent with zero.

We thank P.~Eberhard, G.~Garbarino, B.~Hiesmayr,
A.~Pompili, and F.~Selleri
for insightful discussions,
the KEKB group for excellent operation of the
accelerator, the KEK cryogenics group for efficient solenoid
operations, and the KEK computer group and
the NII for valuable computing and Super-SINET network
support.  We acknowledge support from MEXT and JSPS (Japan);
ARC and DEST (Australia); NSFC and KIP of CAS (China);
DST (India); MOEHRD, KOSEF and KRF (Korea);
KBN (Poland); MIST (Russia); ARRS (Slovenia); SNSF (Switzerland);
NSC and MOE (Taiwan); and DOE (USA).


\bibliography{apssamp}

\begin{thebibliography}{99}
\bibitem{FIT}

For the PS fit, we assign a null deviation when the data falls within the PS range,
otherwise the deviation is computed with respect to the closest boundary:
\APSmax [Eq.~(\ref{eq-APSmax})] or \APSmin [Eq.~(\ref{eq-APSmin})].
\end{thebibliography}

\end{document}